\magnification=1200
\baselineskip=18pt
\rightline{FSU-SCRI-98-16}
\bigskip\bigskip
\centerline{\bf Ginsparg-Wilson relation and the overlap formula}
\bigskip
\centerline{Rajamani Narayanan}
\medskip
\centerline{Supercomputer Computations Research Institute}
\centerline{The Florida State University}
\centerline{Tallahassee, FL 32306-4130}
\vfill
\centerline{\bf Abstract}
\bigskip
The fermionic determinant of a lattice Dirac operator
that obeys the Ginsparg-Wilson relation factorizes into
two factors that are complex conjugate of each other.
Each factor is naturally associated with a single chiral
fermion and can be realized as a overlap of
two many body vacua.
\vfill\eject

The overlap formalism~[1,2] provides a construction of lattice chiral gauge
theories. The chiral determinant on a finite
lattice is realized as an overlap of two
many body vacua~[1] and cannot be written as a path integral over
a finite number of fermi fields on the lattice with a bilinear action.
But the determinant of a massless Dirac fermion in the overlap
formalism can be written as a path integral over a finite number
of fermi fields with a bilinear action~[3] and the lattice Dirac
operator $D$ obeys the Ginsparg-Wilson relation~[4], namely~[5]
$$\gamma_5 D + D\gamma_5 = D \gamma_5 D.\eqno{(1)}$$ 
Recently, it was shown that any lattice fermion action 
$$S_F=\sum_x \bar\psi D \psi \eqno{(2)}$$
with $D$ obeying the Ginsparg-Wilson relation (1) contains
a continuous symmetry that can be viewed as the lattice chiral
symmetry~[6]. 
This would imply that the fermionic determinant, $\det D$, 
should factorize into two pieces with one being the complex conjugate
of the other, if $D$ obeys the Ginsparg-Wilson relation (1). 
Each piece would then be a single chiral determinant on the lattice
associated with a left-handed or right-handed Weyl fermion.
This is indeed the case for the particular form of $D$ considered
in [3] since the operator was obtained starting from the overlap
formalism where the factorization into chiral pieces is built in.
Here I show that the factorization holds as long as $D$ obeys
the Ginsparg-Wilson relation. The two factors are complex conjugates
of each other. Further each factor can be realized as an overlap
of two many body vacua. The proof of factorization presented here
simply amounts to retracing the steps in~[3].

Following Ref.[3,4], we define an operator $\hat H$ through
$$D=1+\gamma_5\hat H.\eqno{(3)}$$
The Ginsparg-Wilson relation (1) reduces to
$$\hat H^2 =1.\eqno{(4)}$$
Therefore all the eigenvalues of $\hat H$ are $\pm 1$. 
Starting in the chiral basis, let 
$$U=\pmatrix {\alpha & \beta \cr \gamma & \delta \cr} \eqno{(4)}$$
be the unitary matrix that diagonalizes $\hat H$ with
$$\hat H \pmatrix {\alpha & \beta \cr \gamma & \delta \cr} = 
\pmatrix {\alpha & -\beta \cr \gamma & -\delta \cr}\eqno{(5)}$$
Under a rotation by $U$,
$$U^\dagger D U = \pmatrix 
{\alpha^\dagger & \gamma^\dagger \cr \beta^\dagger & \delta^\dagger \cr}
\pmatrix{2\alpha & 0 \cr 0 & 2\delta \cr}
\eqno{(6)}$$
If $\hat H$ has an equal number of positive and negative eigenvalues
then $\alpha$ and $\delta$ are square 
matrices.\footnote{${}^\dagger$}
{If $\hat H$ has an unequal number of positive and
negative eigenvalues, $\alpha$ and $\delta$ are not square matrices and
$\det D=0$ implying that the background gauge field has a
non-trivial topology.} 
In this case,
$$\det D = {\det \alpha^\dagger \over \det \delta} \det 2\alpha \det 2\delta
= 2^{2N} \det \alpha \det \alpha^\dagger\eqno{(7)}$$
where $\alpha$ and $\delta$ are assumed to be $N\times N$ matrices.
The first factor on the right side of the first equality in (7)
follows from the fact that $U$ is an unitary matrix.
Eqn. (7) is the factorization of $D$ into two chiral factors one
for each Weyl fermion. 
This factorization comes as no surprise
since $S_F$ in (2) contains a lattice chiral symmetry~[6].

Since the factorization was obtained by simply retracing the steps
in~[3], $\det \alpha$ should be associated with an overlap formula.
To see this, consider the two many body Hamiltonians,
$${\cal H}^- = - a^\dagger \gamma_5 a;\ \ \ \
{\cal H}^+ = - a^\dagger \hat H a \eqno{(8)}$$
with
$a^\dagger$ and $a$ are 
$2N$ fermion creation and annihilation operators
obeying canonical anti-commutation relations.
The matrix $\hat H$ is the $2N\times 2N$ matrix in (3)
and $\gamma_5$ is trivially extended to be a
$2N\times 2N$ matrix. Let $|0\pm>$ be the many body ground
states of ${\cal H}^\pm$. 
The identity in Appendix B of Ref.~[1]
implies that
$$<0-|0+> = \det \alpha \eqno{(8)}$$
with $\alpha$ being the submatrix of $U$ in (4).
Therefore each chiral factor in (7) is equal to an
overlap of two many body vacua. Needless to say the phase 
of the many body ground states plays a crucial role in
the proper construction of chiral gauge theories~[1] but it
does not affect vector gauge theories since $\det D$ is real
and positive and independent of the phase of $\det \alpha$. 

The specific form for $\hat H$ in~[3] is arrived at by writing
$\hat H= {H\over \sqrt{H^2}}$ with $H$ being the Wilson realization
of the continuum $\gamma_5 [\gamma_\mu(\partial_\mu + i A_\mu(x) - m)]$
operator on tha lattice for some $m > 0$. Any other discretization
of the continuum operator can also be used. 
Eqn.~(3) can be thought of as a method to construct a Dirac operator
on the lattice with a lattice chiral symmetry by 
starting from some discretization of
the continuum Dirac operator on the lattice that does not
posses any chiral symmetry. 

We have shown that the determinant of a lattice Dirac operator
that obeys the Ginsparg-Wilson relation factorizes into pieces.
The two pieces are complex conjugate of each other. Each piece
is the determinant of a Weyl fermion and can be thought of as
as an overlap of two many body vacua.

\bigskip
This research was supported by DOE contracts 
DE-FG05-85ER250000 and DE-FG05-96ER40979. The author would
like to thank Robert Edwards, Urs Heller and Herbert Neuberger for
discussions.

\bigskip\bigskip
\centerline{References}
\medskip
\item{[1]} R. Narayanan and H. Neuberger, Nucl. Phys. B412 (1994) 574.
\item{[2]} R. Narayanan and H. Neuberger, Nucl. Phys. B443 (1995) 305.
\item{[3]} H. Neuberger, hep-lat/9707022.
\item{[4]} H. Neuberger, hep-lat/9801031.
\item{[5]} P.H. Ginsparg and K.G. Wilson, Phys. Rev. D25 (1982) 2649.
\item{[6]} M. L\"uscher, hep-lat/9802011.

\end